\documentclass[prl,twocolumn,aps,longbibliography,superscriptaddress,notoc]{revtex4-2}
\usepackage{bm}
\usepackage{graphicx}
\usepackage{amssymb}
\usepackage{amsmath}
\usepackage{eufrak}
\usepackage{color}
\usepackage[utf8]{inputenc}
\usepackage{ulem}
\usepackage[unicode=true,colorlinks=true,citecolor=blue,urlcolor=blue]{hyperref}

\newcommand{\e}{\mathrm{e}}

\let\ifr\i
\renewcommand{\i}{{\rm i}}
\renewcommand{\d}{\mathrm d}
\renewcommand{\emph}{\textit}
\newcommand{\braket}[1]{\left\langle #1 \right\rangle}

\newcommand{\addDima}[1]{{#1}}

\usepackage{chngcntr}
\newcommand{\enquote}{}

\newcommand{\nix}[1]{}
\let\oldsec\section
\renewcommand{\section}[1]{\textit{#1}---}

\begin{document}

\title{Spin light emitting diode based on exciton fine structure tuning in quantum dots}

\author{A.~V.~Shumilin}
\affiliation{Ioffe Institute, 194021 St. Petersburg, Russia}
\affiliation{Complex Matter Department, Jozef Stefan Institute, Jamova 39, SI-1000 Ljubljana, Slovenia}

\author{T.~S.~Shamirzaev}
\affiliation{Rzhanov Institute of Semiconductor Physics, Siberian Branch of the Russian Academy of Sciences, 630090 Novosibirsk, Russia}

\author{D.~S.~Smirnov}
\affiliation{Ioffe Institute, 194021 St. Petersburg, Russia}
\email[Electronic address: ]{smirnov@mail.ioffe.ru}

\begin{abstract}
  We propose a concept of quantum dot based light emitting diode that produces circularly polarized light due to the tuning of the exciton fine structure by magnetic field and electron nuclear hyperfine interaction. The device operates under injection of electrons and holes from nonmagnetic contacts in a small field of the order of milliteslas. Its size can be parametrically smaller than the light wavelength, and circular polarization degree of electroluminescence can reach 100\%. The proposed concept is compatible with the micropillar cavities, which allows for the deterministic electrical generation of single circularly polarized photons.
\end{abstract}

\maketitle{}

\section{Introduction}The compact sources of cicularly polarized light are strongly anticipated for the opto-spintronic applications in general and for the quantum information processing in particular~\cite{Atature2018,Huebener2020}. The circularly polarized light can be obtained using the optical activity of bulk crystals, heterostructures~\cite{Kotova2016,Poshakinskiy2018,Michl2022}, or organic molecules~\cite{Ha2007,Pu2021}.

Much more compact devices exploit chiral vacuum electromagnetic modes, which can be created in chiral photonic crystals, using chiral metasurfaces~\cite{Konishi2011,Shitrit2013,Lobanov2015,Li2018a}, or exploiting a chiral design of heterostructures~\cite{Lin331,Spitzer2018,Yin2020}.


Even more compact and technologically simple sources of circularly polarized light are based on the spin polarization of the charge carriers involved in the light emission. For example, absorption of circularly polarized light leads to optical spin orientation and reemission of circularly polarized light provided the spin relaxation time is long enough~\cite{ivchenko05a}. For practical applications, the electrical spin injection is more favorable, and such sources of circularly polarized light are widely known as spin-light emitting diodes (LEDs)~\cite{Holub2007}. In spin-LEDs, the spin is injected to single or multiple quantum wells or quantum dots (QDs) from ferromagnetic contacts~\cite{Ohno99_2,Fiederling99,Chye2002,Jiang2005,Hamaya2009,Chen2014,Nishizawa2017,Lindemann2019} or chiral environment, such as organic molecules~\cite{Yang2013,Kim2021}.

We put forward an alternative concept of spin-LEDs \addDima{that exploits} the fine structure tuning of excitons in semiconductor QDs by magnetic field. In it,
no ferromagnetic contacts are needed, the sense of the circular polarization of the electroluminescence (EL) is determined by a tiny magnetic field of the order of ten millitesla, and the size of the device can be much smaller than the light wavelength.

Tuning of the exciton fine structure by magnetic field can lead to the so-called dynamic electron spin polarization effect, i.e. appearance of a large electron spin polarization in nonequilibrium conditions due to the electron-nuclear flip-flops, which can be used to produce circularly polarized light. This effect was first described and observed in momentum indirect type-I (In,Al)As/AlAs QDs~\cite{PhysRevLett.125.156801}, which show long spin relaxation times up to milliseconds at zero magnetic field. Later dynamic electron spin polarization effect was described theoretically~\cite{PhysRevB.104.L241401} and observed experimentally~\cite{Wang2022} for moir\'e quantum dots in twisted transition metal dichalcogenides heterobilayers. It was also shown to take place under electric current flow in organic semiconductor molecules~\cite{Shumilin2022}, which are often used \addDima{for spin}-LEDs~\cite{Forrest2004,Wei2018}.

We propose \addDima{a} concept of spin-LEDs \addDima{that is} based on the dynamic electron spin polarization effect in semiconductor QDs. Such structures are compatible with the micropillar cavities~\cite{Somaschi2016,Unsleber2016,Galimov2021,Tomm2021}, which can be used enhance the emission directionality and produce single photons~\cite{Bennett2005,Heindel2010}. In combination, this allows for the deterministic generation of single circularly polarized photons by electric pulses and promises to find an important application in quantum communications and photon-based computations.

\section{Model}
\begin{figure}
  \centering
  \includegraphics[width=0.95\linewidth]{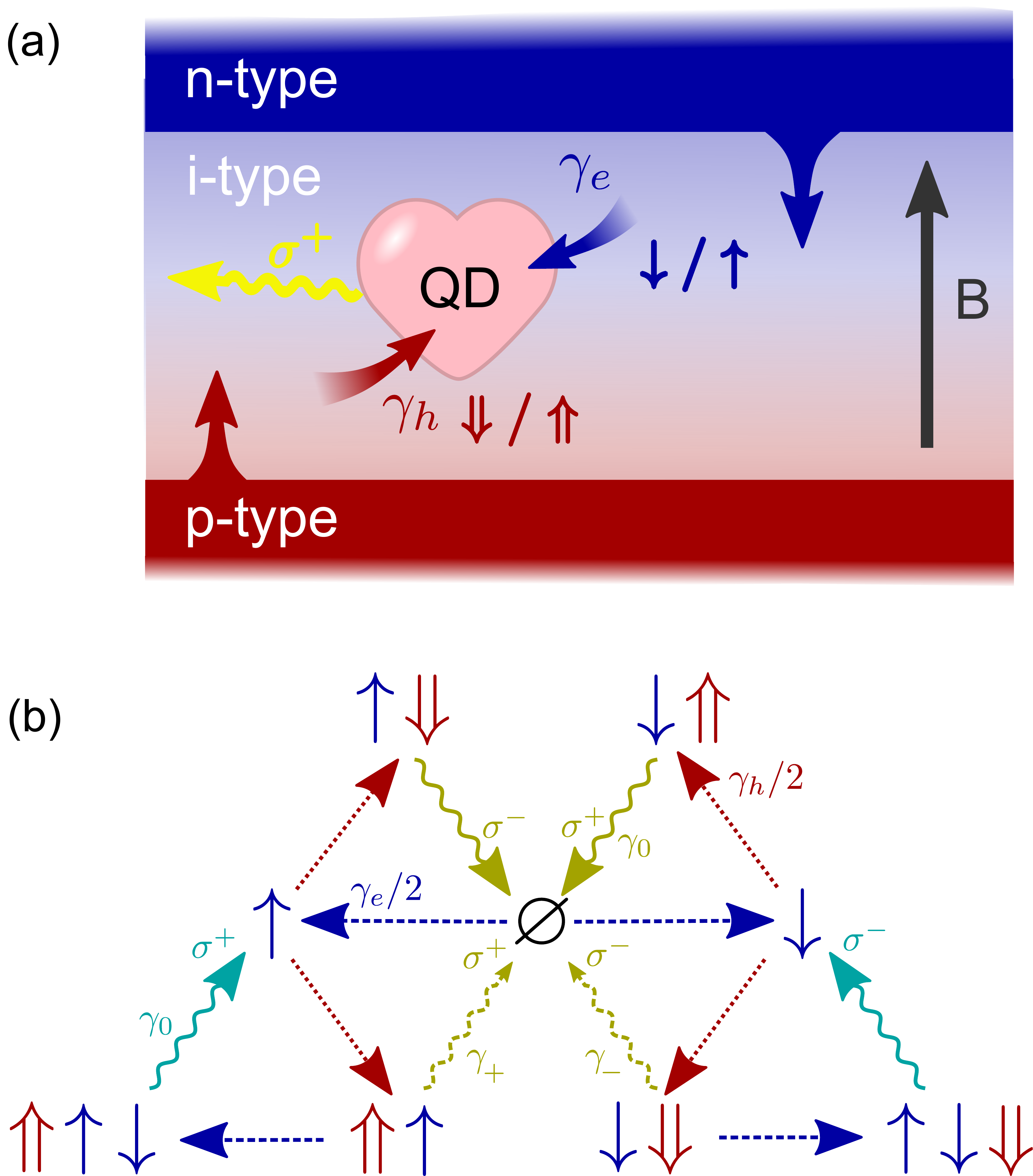}
  \caption{(a) Electrical injection of electrons (blue) and holes (red) to a QD (heart of the device), which emits circularly polarized light (yellow). (b) Transitions between different QD states. Blue and red solid arrows denote electron and hole spins, blue dashed and red dotted arrows denote capture of electrons and holes, respectively, yellow and dark cyan wavy arrows are the exciton and trion radiative recombinations, which obey the optical selection rules. 
  }
  \label{fig:system}
\end{figure}

The device under consideration basically represents a QD based LED in external magnetic field with electrical injection of electrons and holes from $n$-type and $p$-type contacts to intermediate intrinsic layer with self-assembled QDs, see Fig.~\ref{fig:system}(a). The QDs are assumed to be made of GaAs-type semiconductors and to be small enough so that only the first size quantized levels of electron and hole in QDs can be considered (and we neglect the possible valley degeneracy). We denote the electron and hole capture rates to an empty QD as $\gamma_e$ and $\gamma_h$, respectively. The single particle states are Kramers degenerate with respect to the electron spin $S_z=\pm1/2$ and the heavy hole spin $J_z=\pm3/2$. When a QD captures both an electron and a hole, an exciton forms. It can recombine radiatively with the rate $\gamma_0$ in the bright spin state, according to the optical selection rules~\cite{ivchenko05a}: $S_z=\mp1/2$ electron can recombine with $J_z=\pm3/2$ hole only emitting circularly polarized $\sigma^\pm$ photon, respectively.

We consider the limit of
\begin{equation}
  \label{eq:rates}
  \gamma_0\gg\gamma_e\gg\gamma_h,
\end{equation}
when the radiative recombination rate of bright excitons is the fastest and the hole capture rate is the slowest in the system, while the electron capture rate is in between them. We also assume the strong Coulomb interaction, which forbids occupancy of a QD by two electrons without a hole. The possible transitions between QD states are shown in Fig.~\ref{fig:system}(b): Starting from the empty QD state ($\varnothing$), first, an electron with spin-up or -down is captured by the QD (blue dashed arrows) because of the assumptions of much slower hole capture rate. Then due to the Coulomb blockade only a hole can be captured with the rate $\gamma_h$ (red dotted arrows). Afterwards, if the bright exciton is formed, it quickly radiatively recombines (yellow solid wavy arrows). If the dark exciton with parallel electron and hole spins is formed, there are two options: (i) It can recombine radiatively [yellow dashed wavy arrows in Fig.~\ref{fig:system}(b)] due to the mixing of bright and dark excitonic states by the hyperfine interaction. The corresponding recombination rate $\gamma_\pm$ is small ($\sim\gamma_e\ll\gamma_0$) and depends on the spin of the hole in the exciton $J_z=\pm3/2$. (ii) A second electron with the opposite spin can be captured with the rate $\gamma_e/2$. In the latter case, the singlet trion forms and quickly recombines radiatively with the rate $\gamma_0$. Then a single electron remains in the QD (dark cyan wavy arrows). The interplay between these two options produces the circular polarization of the emitted light, as we show below.

The mixing of bright and dark excitons is provided mainly by the electron hyperfine interaction with the spins of the host lattice nuclei, while the hyperfine interaction of holes is much weaker~\cite{book_Glazov}. \addDima{However, the mixing is} suppressed by the electron-hole short-range exchange interaction, which leads to the energy splitting between bright and dark exciton states (we neglect the smaller anisotropic long-range electron hole exchange interaction). This fine structure can be tuned by application of external longitudinal magnetic field $\bm B$ in the Faraday geometry. All these interactions conserve the heavy hole spin $J_z$, so the exciton spin dynamics can be described by the Bloch equation for the electron spin $\bm S$~\cite{shamirzaev2021dynamic}:
\begin{equation}
  \frac{\d\bm S}{\d t}=\frac{g_e\mu_B}{\hbar}\bm B_{\rm tot}\times\bm S.
\end{equation}
Here $g_e$ is the electron $g$ factor, $\mu_B$ is the Bohr magneton, and
\begin{equation}
  \label{eq:Btot}
  \bm B_{\rm tot}=\bm B+\bm B_n+\frac{2J_z}{3}B_{\rm exch}\bm e_z
\end{equation}
is the total effective magnetic field, which includes the external field $\bm B$, the exchange field $\pm B_{\rm exch}$ along the $z$ axis~\cite{Astakhov07}, and the Overhauser field $\bm B_n$ of randomly oriented nuclear spins. The nuclear spin dynamics is slow and can be neglected at the time scale of the electron spin dynamics. So the nuclear spin fluctuations effectively produce random ``frozen'' Overhauser field with the Gaussian probability distribution function $\propto\exp(-B_n^2/\Delta_B^2)$, where $\Delta_B$ characterizes the dispersion. In the limit of fast electron spin precession we obtain the recombination rates of the dark excitons with $J_z=\pm3/2$ caused by the mixing with the bright excitons~\cite{supp}:
\begin{equation}
  \gamma_\pm=\gamma_0\frac{B_{n,x}^2+B_{n,y}^2}{(B\pm B_{\rm exch})^2}.
  \label{eq:gamma_pm}
\end{equation}
The recombination takes place due to the transverse components of the Overhauser field and its rate is inversely proportional to the squared splitting between the bright and dark excitons \addDima{$\propto B\pm B_{\rm exch}$. We note that the mixing can be provided by the tilted external magnetic field as well.}

\begin{figure}
  \includegraphics[width=\linewidth]{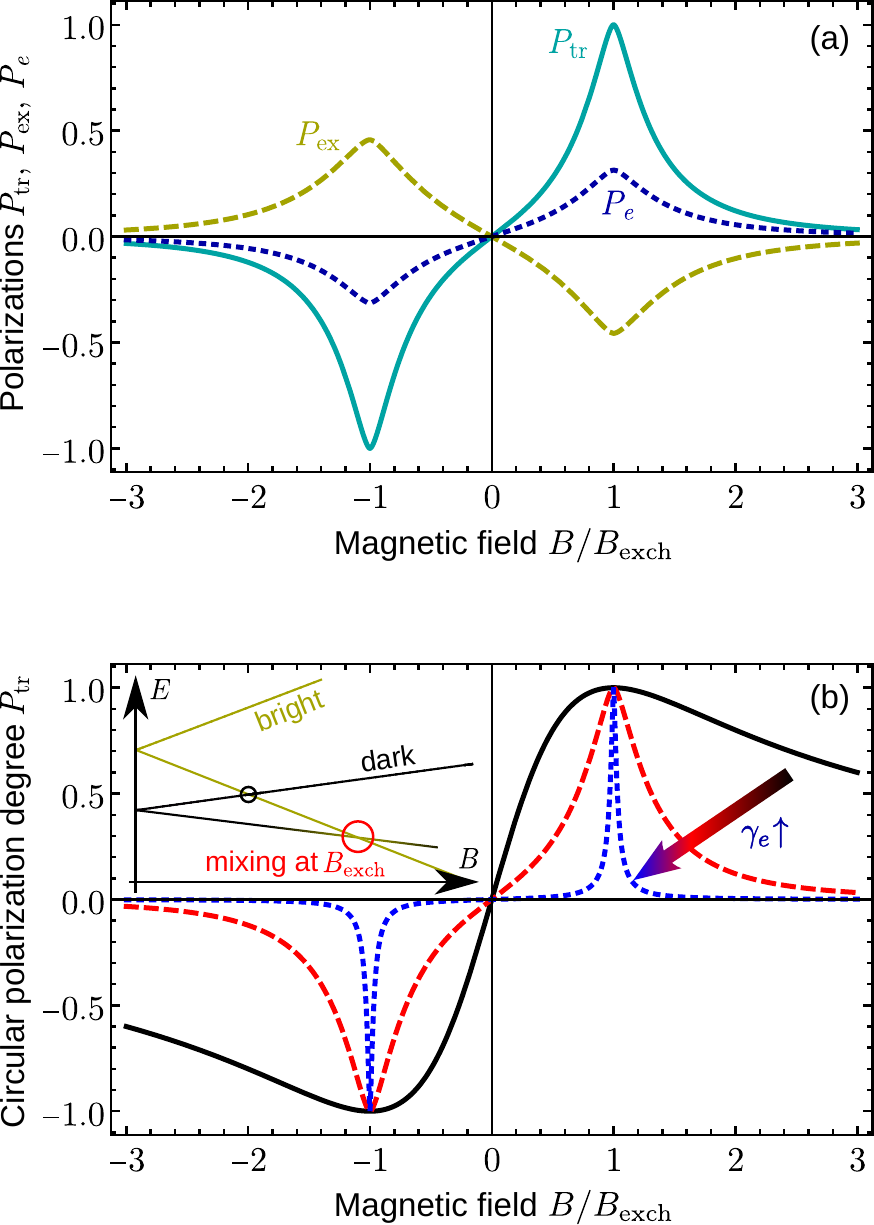}
  \caption{\label{fig:main}
    (a) Polarizations of trion EL (solid dark cyan line), exciton EL (yellow dashed line), and an electron (blue dotted line) for $\mathcal R=10$. (b) Polarization of trion EL for increaseing electron injection rate: $\mathcal R\to 0$ (black solid line), $\mathcal R=10$ (red dashed line), and $\mathcal R=10^3$ (blue dotted line). The inset shows the \addDima{splitting of bright and dark exciton states in magnetic field. At the level crossing shown by the red circle $B=B_{\rm exch}$ and} the polarizations are the largest.}
\end{figure}

Under the assumptions~\eqref{eq:rates}, the QD is occupied most of the time with a single electron, so its state can be characterized by the two occupancies $p_{\uparrow/\downarrow}$ of spin-up/down states. From Fig.~\ref{fig:system}(b) one can see that the rates of the trion recombination with emission of $\sigma^\pm$ photons are given by~\cite{supp}
\begin{equation}
  I_\pm^{\rm tr}=\frac{p_{\uparrow/\downarrow}\gamma_e\gamma_h/4}{\gamma_e/2+\gamma_\pm}.
  \label{eq:Ipm}
\end{equation}
At the same time, the rates of $\sigma^\pm$ photon emission by excitons are $I_\pm^{\rm ex}=\gamma_h/2-I_\pm^{\rm tr}$, so the total EL is unpolarized. 

The kinetic equations for the occupancies read~\cite{supp}
\begin{equation}
  \frac{\d p_{\uparrow/\downarrow}}{\d t}=\pm\frac{1}{2}\left[\gamma_h(p_{\downarrow}-p_{\uparrow})+I_+^{\rm tr}-I_-^{\rm tr}\right].
  \label{eq:kinetic}
\end{equation}
From these equations we find the intensities of the exciton and trion EL in the steady state $\braket{I^{\rm ex/tr}}=\braket{I_+^{\rm ex/tr}+I_-^{\rm ex/tr}}$, their polarizations $P_{\rm ex/tr}=\braket{I_+^{\rm ex/tr}-I_-^{\rm ex/tr}}/ \braket{ I^{\rm ex/tr} }$, and the electron spin polarization $P_e=\braket{p_{\uparrow}-p_{\downarrow}}$, where the angular brackets denote the averaging over the random Overhauser field. These averages can be calculated analytically, but the results are very cumbersome. Generally, the polarizations are determined by the two dimensionless parameters $B/B_{\rm exch}$ and $\mathcal R=\gamma_e B_{\rm exch}^2/(\gamma_0\Delta_B^2)$. These parameters can be tuned by the external magnetic field and the electron injection rate, respectively.

We show the polarization degrees of trion and exciton EL and of the electron spin in Fig.~\ref{fig:main}(a) as functions of the magnetic field. These functions are odd in agreement with the time reversal symmetry and reach the largest values at $|B|=B_{\rm exch}$ (we assume $B_{\rm exch}>0$, which corresponds to the negative electron $g$ factor). This can be understood from the inset in Fig.~\ref{fig:main}(b), which shows the splitting of the exciton levels in magnetic field. At $B=\pm B_{\rm exch}$ the levels of a bright and a dark exciton with the same hole spin cross (red circle in the inset), so their mixing due to the hyperfine interaction is the strongest. At this field, the radiative recombination rate of one of the dark excitons has a resonance, while that of another is small, see Eq.~\eqref{eq:gamma_pm}. This leads to the strong difference between the transitions in the left and right sides of Fig.~\ref{fig:system}(b), so the polarizations of electron, exciton and trion EL are the largest. Note that there is also another level crossing shown by a small black circle in the inset in Fig.~\ref{fig:main}(b), but the corresponding mixing between bright and dark states requires hole spin flips, which we neglect.

In our model, the electron spin polarization can not exceed $1/3$, the exciton EL polarization is generally smaller than $1/2$, but the trion EL polarization can reach 100\%. So for the rest of the paper we focus on the latter. We note that it is well separated from the exciton EL by the trion binding energy, which allows one to easily select it experimentally. Fig.~\ref{fig:main}(b) shows the dependence of the trion EL polarization on the magnetic field, which can have qualitatively different shape. Provided $\mathcal R\ll 1$, we obtain $P_{\rm tr}=2BB_{\rm exch}/(B^2+B_{\rm exch}^2)$~\cite{supp}. So this dependence is broad and slowly decays with increase of the magnetic field in this limit, as shown by the black solid line in Fig.~\ref{fig:main}(b). With increase of $\mathcal R$, the narrow \addDima{resonances} appear in the polarization at $B=\pm \addDima{B_{\rm exch}}$, as shown by the red dashed and blue dotted lines. In the limit of $\mathcal R\gg 1$, the dynamic electron spin polarization effect basically reduces to the magnetic circular polarization of exciton photoluminescence effect~\cite{Ivchenko2018}.

Notably, in all cases, the degree of circular polarization reaches $100\%$ in maximum, which underlines universality of our concept for spin-LEDs based on the exciton fine structure tuning.

\section{Single photon source}
\begin{figure*}
  \includegraphics[width=0.9\linewidth]{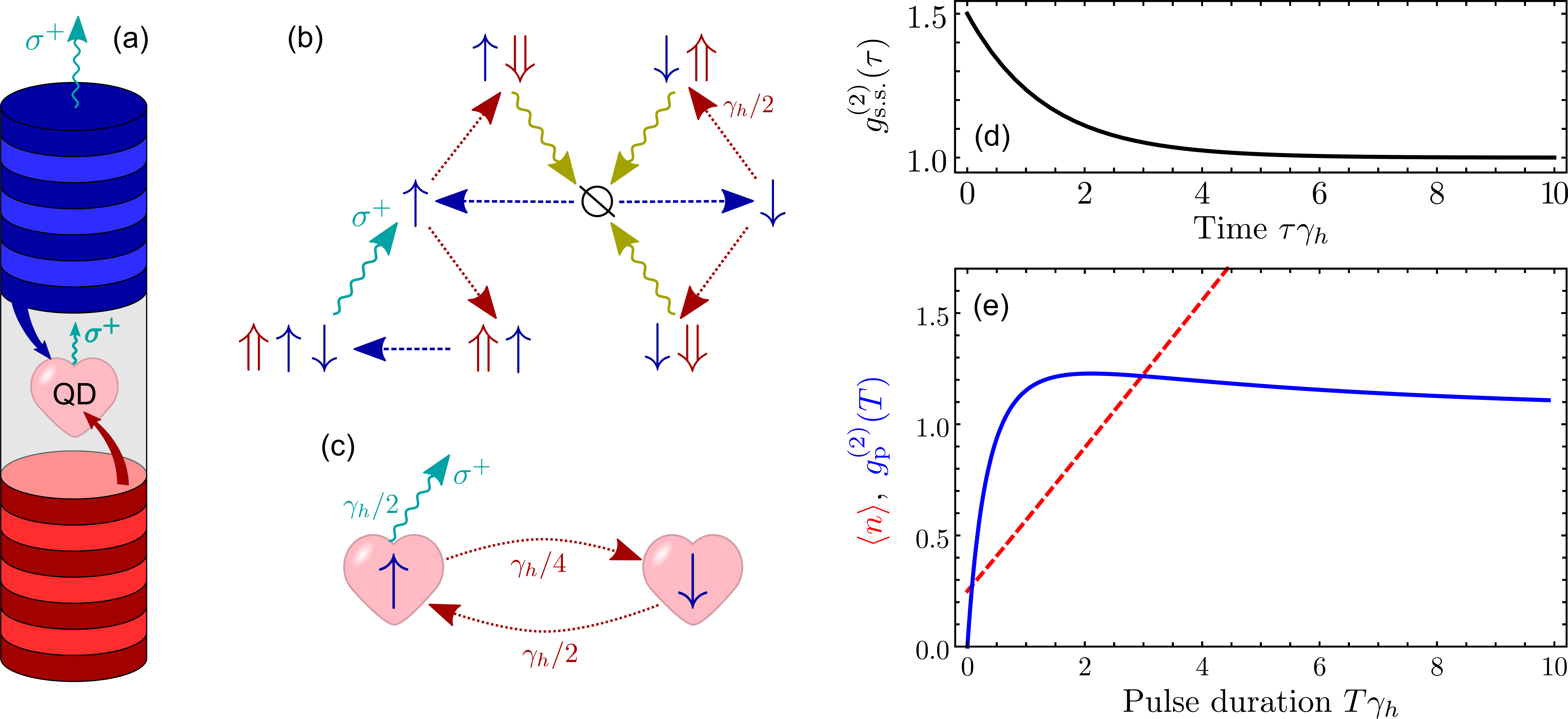}
  \caption{\label{fig:1photon}
    (a) Sketch of a QD micropillar cavity with electron and hole injection. (b) Simplified scheme of transitions between QD states from Fig.~\ref{fig:system}(b) under resonant condition $B=B_{\rm exch}$. (c) Effective transitions between electron spin-up and spin-down states of the QD and their rates obtained from (b) in the limit~\eqref{eq:rates}. (d) Second order photon correlation function in the steady state~\cite{supp}. (e) EL intensity (red dashed line) and second order photon correlation function (blue solid line) under pulsed excitation as functions of the pulse duration~\cite{supp}.}
\end{figure*}

The self-assembled QDs can be easily embedded in micropillar cavities to enhance light-matter interaction and to increase directionality of the emission~\cite{Somaschi2016,Unsleber2016,Galimov2021,Tomm2021}. A sketch of the device is shown in Fig.~\ref{fig:1photon}(a). Electrons and holes can be injected electrically into the single QD and an additional contact to intrinsic layer can allow one to change the rates $\gamma_e$ and $\gamma_h$ independently. The electron hole recombination at the level crossing ($B=B_{\rm exch}$) \addDima{produces} circularly polarized light, and one can tune the microcavity resonance frequency to select the emission at the trion resonance frequency and suppress another frequencies.

The counting statistics of the emitted photons can be described on the basis of Eqs.~\eqref{eq:Ipm} and~\eqref{eq:kinetic}. The level crossing is described by $\gamma_-\gg\gamma_e\gg\gamma_+$, and the corresponding transitions are shown in Fig.~\ref{fig:1photon}(b). The simplified kinetic equations read
  \begin{equation}
    \frac{\d p_{\uparrow/\downarrow}}{\d t}=\pm\gamma_h\left(\frac{1}{2}p_\downarrow-\frac{1}{4}p_\uparrow\right),
  \label{eq:kin_lim_main}
\end{equation}
see Fig.~\ref{fig:1photon}(c). These equations are valid at the slowest time scale $\sim1/\gamma_h$. After the trion recombination with emission of $\sigma^+$ photon, the system is left in the electron spin-up state. The next photon can not be emitted immediately after this, so the emitted photons are antibunched at the time scale $\sim1/\gamma_0$. However, at the longer time scale, the increased occupation of the electron spin-up state increases the probability of the second $\sigma^+$ photon emission relative to the steady state. This leads to the bunching of photons emitted from the device at the time scale $\sim1/\gamma_h$. The second order photon  correlation function $g^{(2)}_{s.s.}(\tau) = \langle I^{\rm tr}(t) I^{\rm tr}(t + \tau) \rangle/\langle I^{\rm tr}(t)\rangle^2$ is shown in Fig.~\ref{fig:1photon}(d) and decreases from $1.5$ to $1$. The antibunching at the short time scale is not shown.

The operation of our device as a single-photon source requires the antibunching at the long time scale, instead of the bunching. This can be achieved by applying electrical pulses. Thus, we consider a small constant hole capture rate $\gamma_h$ and a pulsed electron capture rate $\gamma_e$ during the pulse duration $T$. In this case, the QD is initially charged with a single hole with the unpolarized spin. After beginning of the pulse, satisfying Eq.~\eqref{eq:rates}, the EL appears. A single photon can be emitted immediately after injection of an electron. However, the second photon can not be emitted before another hole is captured with the slow rate $\gamma_h$. As a result, the electric pulses produce strongly antibunched EL when $T\lesssim1/\gamma_h$, even if their length is much longer than the radiative recombination time and the inverse electron capture rate.

The photon statistics can be described on the same basis of Eq.~\eqref{eq:kin_lim_main}. The average number of emitted photons per pulse $\braket{n}$ is shown in Fig.~\ref{fig:1photon}(e) by the red dashed line as a function of the pulse duration. It starts from $1/4$, because we neglect the time scales $\sim1/\gamma_e$, so the fast single electron injection leads to the photon emission at the trion frequency with the probability $1/4$, as can be seen from Fig.~\ref{fig:1photon}(b). Average number of photons grows almost linearly at the time scale $\sim1/\gamma_h$~\cite{supp}.

The photon correlation function in the pulsed regime is given by
\begin{equation}
  g_p^{(2)}(T)=\frac{\braket{n(n-1)}}{\braket{n}^2},
\end{equation}
where the angular brackets denote the quantum statistical averaging. It is shown in Fig.~\ref{fig:1photon}(e) by the blue solid line as a function of the electrical pulse duration (an explicit expression for it is given in the supplemental material~\cite{supp}). One can see a perfect antibunching $g^{(2)}_p(0)=0$ in the limit of the short pulses. Most importantly, the EL remains antibunched ($g^{(2)}_p(T)<1$) even for the long pulses with $T<0.6/\gamma_h$ in contrast to the steady state EL. With further increase of $T$ the EL becomes bunched ($g^{(2)}_p(T)>1$) similar to the steady state, and finally $g^{(2)}_p$ tends to one at $T\gg1/\gamma_h$, because the average number of emitted photons per pulse $\braket{n}$ becomes large.

\section{Discussion}The performance of a single photon source is characterized by its brightness and purity of single photons~\cite{Bennett2005,Heindel2010}. The former can be strongly increased by using zero dimensional microcavities~\cite{Tomm2021}, and both depend on the product of the electrical pulse duration and hole capture rate: $\braket{n}\approx(1+1.25T\gamma_h)/4$ and $g_p^{(2)}(T)\approx4T\gamma_h$. From these expressions one can see that the product $T\gamma_h$ should be kept as small as possible even at the expense of the reduced intensity. In practice, the minimal value of this parameter can be limited by the minimal electrical pulse duration and by the condition that $\gamma_h$ should be much larger than the electron leakage current when the electron injection is switched off.


The dynamic electron spin polarization effect has an important advantage of temperature independence at low temperatures, provided the electron spin relaxation is driven by the hyperfine interaction~\cite{PhysRevLett.125.156801}. In addition, it allows for the generation of circularly polarized light \addDima{by QDs} with the sizes of the order of 10~nm, which can be directly integrated into the optical fibers.

We also stress that in our concept of spin-LEDs no magnetic components are required, and the circularly polarized EL is produced in a small magnetic field $B=B_{\rm exch}$. The exact value of the exchange magnetic field $B_{\rm exch}$ strongly depends on the material of the QD and its band structure. In direct quantum dots, it is typically as large as a few Tesla~\cite{book_Glazov}, but for the indirect excitons in real or in momentum space it is exponentially suppressed by \addDima{a} small overlap of the electron and hole wave functions. For example, in InAlAs QDs the exchange field was measured to be $5.5$~mT~\cite{PhysRevLett.125.156801,shamirzaev2021dynamic}, but it can be even smaller~\cite{nano13040729}. Such small magnetic fields can be easily generated and switched, which allows one to control the degree and the sense of EL circular polarization.

\textit{In conclusion}, we have put forward a concept of spin-LEDs based on the effect of dynamic electron spin polarization due to the exciton fine structure tuning by magnetic field and electron hyperfine interaction. In this concept, the spin-LEDs do not require magnetic or chiral elements and can produce completely circularly polarized light, having a size smaller than the wavelength. 
By embedding a single QD in a zero dimensional microcavity one can get a device, which on demand produces single circularly polarized photons with variable helicity.

  We acknowledge the Foundation for the Advancement of Theoretical Physics and Mathematics ``BASIS''. Description of the spin-LEDs by A.V.S. was supported by the Russian Science Foundation grant No. 22-12-00125. The development of the device structure by T.S.Sh was supported by the Russian Science Foundation grant No. 22-12-00022. The description of single photon EL by D.S.S. was supported by the Russian Science Foundation grant No. 23-12-00142.


\renewcommand{\i}{\ifr}
\let\oldaddcontentsline\addcontentsline
\renewcommand{\addcontentsline}[3]{}

%

\let\addcontentsline\oldaddcontentsline
\makeatletter
\renewcommand\tableofcontents{%
    \@starttoc{toc}%
}
\makeatother
\renewcommand{\i}{{\rm i}}

\appendix

\onecolumngrid
\vspace{\columnsep}
\begin{center}
\makeatletter
{\large\bf{Supplemental Material to\\``\@title''}}
\makeatother
\end{center}
\vspace{\columnsep}

The Supplementary Material includes the following topics:

\hypersetup{linktoc=page}
\tableofcontents
\vspace{\columnsep}
\twocolumngrid

\counterwithin{figure}{section}
\renewcommand{\section}[1]{\oldsec{#1}}
\renewcommand{\thepage}{S\arabic{page}}
\renewcommand{\theequation}{S\arabic{equation}}
\renewcommand{\thefigure}{S\arabic{figure}}

\setcounter{page}{1}
\setcounter{section}{0}
\setcounter{equation}{0}
\setcounter{figure}{0}

\section{S1. Derivation of kinetic equations}
\label{sec:S1}

\begin{figure}[b]
  \includegraphics[width=\linewidth]{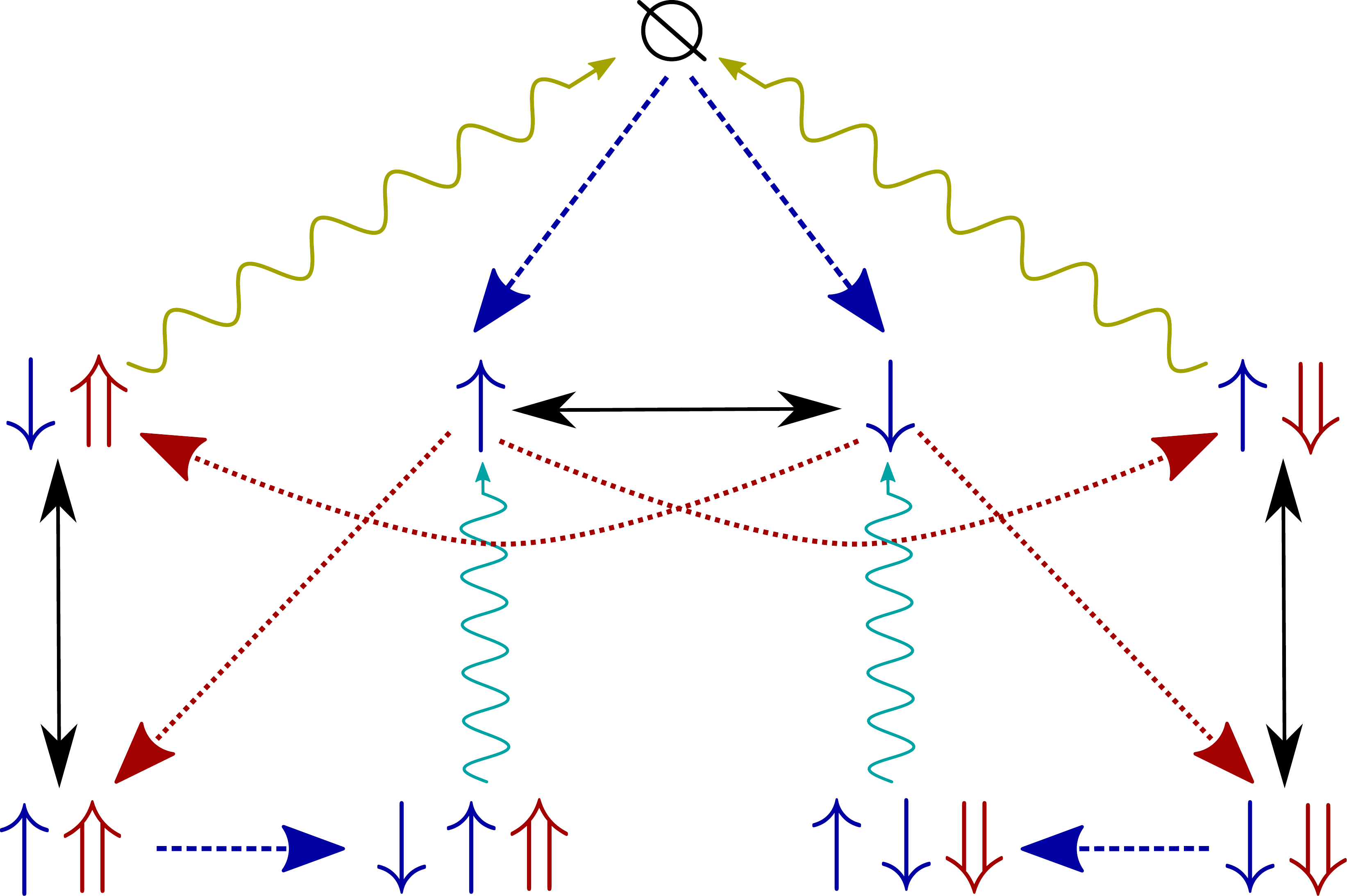}
  \caption{\label{fig:all_transitions}
    Transitions between different QD states, see the \addDima{text for} the notations.
  }
\end{figure}

We derive the model of the main text as illustrated in Fig.~\ref{fig:all_transitions}. An empty QD (\addDima{$\varnothing$}) can capture either spin-up or spin-down electron with the rate $\gamma_e$ (blue dashed arrows) because of the assumption of much slower hole capture rate, Eq.~\eqref{eq:rates} in the main text. Then the Coulomb repulsion blocks capture of the second electron by the QD, so it can capture either spin-up or spin-down hole with the rate $\gamma_h$ (red dotted arrows). If a bright exciton is formed, it quickly radiatively recombines with the rate $\gamma_0$ (wavy dark yellow arrows). If a dark exciton forms, it survives until the second electron with the opposite spin is captured with the rate $\gamma_e/2$. Then this electron recombines with a hole and a single electron remains in the QD.

In addition to the above inelastic processes, we account for the coherent spin dynamics in the QD taking into account external longitudinal magnetic field $\bm B$ applied along the $z$ axis, and random Overhauser field $\bm B_n$, which acts only on electrons because of the suppressed hyperfine interaction for holes~\cite{book_Glazov}. This, generally conserves the hole spin $J_z$, and leads to the precession of the electron spin $\bm S$ in singly charged QDs and in QDs with excitons (black arrows in Fig.~\ref{fig:all_transitions}):
\begin{equation}
  \frac{\d\bm S}{\d t}=\frac{g_e\mu_B}{\hbar}\bm B_{\rm tot}\times\bm S.
\end{equation}
Here $g_e$ is the electron $g$ factor, $\mu_B$ is the Bohr magneton, and $\bm B_{\rm tot}$ is the total effective magnetic field given by Eq.~\eqref{eq:Btot} in the main text, where $J_z=\pm3/2$ for excitons, and $J_z$ should be set to zero for a single electron in the QD. We neglect the slow nuclear spin dynamics and describe the random Overhauser field by the Gaussian distribution function
\begin{equation}
  \mathcal F(\bm B_n)=\frac{1}{\left(\sqrt{\pi}\Delta_B\right)^3}\exp\left(-\frac{B_n^2}{\Delta_B^2}\right).
\end{equation}
We also neglect the anisotropic long-range exchange interaction, which in practice should be weaker than the hyperfine interaction. Finally, we assume that $B_{\rm exch}\gg\Delta_B$ and the electron spin precession is faster than all the inelastic processes, i.e. $g_e \mu_B \Delta_B/\hbar\gg\gamma_0$.

The fast spin precession in the field $\bm B_{\rm tot}$ leads to the mixing of dark and bright excitons, so the dark exciton eigenstates become pseudo dark~\cite{shamirzaev2021dynamic} and obtain a finite radiative recombination rate
\begin{equation}
  \label{gammaPM}
  \gamma_{\pm} = \gamma_0\frac{1-\cos\theta_{\pm}}{2}
\end{equation}
Here, the subscript $\pm$ corresponds to $J_z=\pm 3/2$ and $\theta_{\pm}$ is the angle between $\bm B_{\rm tot}$ and the $z$ axis.

A QD with a pseudo dark exciton can capture a second electron with the rate $\gamma_e/2$ obeying the Pauli principle. This leads to the fast recombination of the forming trion. Therefore, the probability for a QD with a pseudo dark exciton to eventually emit a photon at the trion frequency is $\gamma_e/(\gamma_e + 2\gamma_\pm)$. After the trion recombination, the QD is occupied with the single electron with spin parallel to the hole spin in the trion. This spin precesses in the field $\bm B+\bm B_n$, which leads to the two consequences. First, the electron spin can flip after the hole absorbtion due to the difference between $\bm B+\bm B_n$ and $\bm B_{\rm tot}$ with the probability $(1-\cos\beta_{\pm})/2$ where $\beta_{\pm}$ is the angle between effective fields with and without the hole. It results in the following emission rates at the trion frequency
\begin{equation}
  \label{ItrFull}
 I_\pm^{\rm tr} = \frac{\gamma_h\gamma_e/4}{\gamma_e/2 + \gamma_\pm}
  \left(p_{\uparrow/\downarrow} \frac{1 +   \cos\beta_\pm}{2} +
    p_{\downarrow/\uparrow} \frac{1 - \cos \beta_\pm}{2}  \right),
\end{equation}
where $p_\uparrow$ and $p_\downarrow$ are occupancies of the QD with spin-up and spin-down electrons, respectively. Second, after the trion recombination, the electron spin also flips with probability $(1-\cos\alpha)/2$ where $\alpha$ is the angle between the $z$ axis and the axis of $\bm B+\bm B_n$. This leads to the kinetic equation
\begin{equation}\label{s_kin}
  \frac{\d p_{\uparrow/\downarrow}}{\d t} = \pm \frac{1}{2}\left[ \gamma_h \left(p_\downarrow - p_\uparrow \right) + \cos\alpha \left(I_+^{\rm tr} - I_-^{\rm tr}\right) \right].
\end{equation}

The model is significantly simplified by the assumption of $B_{\rm exch} \gg \Delta_B$. In this limit, $\bm B+\bm B_n$ is directed along the $z$ axis leading to $\alpha  = 0$ and $\beta_{\pm} = \theta_{\pm}$ in all the relevant cases. Also because of $\gamma_0 \gg \gamma_e$ the trions can appear only when the mixing between dark and bright excitons is weak, $\theta_{\pm} \ll 1$. This allows us co consider $\cos\beta_{\pm} = 1$. As a result, we obtain Eqs.~\eqref{eq:Ipm} and~\eqref{eq:kinetic} of the main text with $\gamma_{\pm}$ given by Eq.~\eqref{gammaPM}, which reduces to Eq.~\eqref{eq:gamma_pm} of the main text in the limit of $\Delta_B \ll B_{\rm exch}$.

After averaging of the occupancies and the intensities in the steady state over the random Overhauser field, we obtain
\begin{multline}
  \braket{I_\pm^{\rm tr}}=\frac{(1\pm b)^2\mathcal R\gamma_h}{16}\sum_{\sigma=\pm}\sigma\left(\frac{1\pm 4b+b^2}{\sqrt{1+34b^2+b^4}}-\sigma\right)\\
  \times\exp\left(a_\sigma\right){\rm Ei}\left(-a_\sigma\right),
\end{multline}
where $b=B/B_{\rm exch}$,
\begin{equation}
  a_\sigma=\frac{\mathcal R}{8}\left(3+3b^2-\sigma\sqrt{1+34b^2+b^4}\right),
\end{equation}
and ${\rm Ei}(x)$ is the exponential integral function. From this we calculate $P_{\rm tr}$, $P_{\rm ex}$, and $P_e$ in the main text.

In particular, in the limit of $\mathcal R\to 0$, we obtain
\begin{equation}
  P_{\rm tr}=\frac{2BB_{\rm exch}}{B_{\rm exhc}^2+B^2}.
\end{equation}
In the opposite limit of $\mathcal R\to\infty$ we find
\begin{equation}
  P_{\rm tr}=\sum_{\sigma=\pm}\sigma\left[\frac{18}{6+\delta_\sigma^2\e^{\delta_\sigma^2/3}{\rm Ei}(-\delta_\sigma^2/3)}-2\right],
\end{equation}
where $\delta_\sigma=(B+\sigma B_{\rm exch})\sqrt{\mathcal R}$. Curiously we find that this expression can be approximated as
\begin{equation}
  P_{\rm tr}\approx\sum_{\sigma=\pm}\frac{\sigma}{1+|\delta_\sigma|^{3/2}}
\end{equation}
with the accuracy better than $0.02$.

\section{S2. Calculation of $g^{(2)}$ function at level crossing}

In this section, we consider the limit of resonant magnetic field $B\approx B_{\text{exch}}$, when levels of bright and dark excitons cross. In this limit, we obtain from Eq.~\eqref{eq:gamma_pm} $\gamma_-\gg\gamma_e$ and $\gamma_+\ll\gamma_e$. Then from Eq.~\eqref{eq:Ipm} in the main text we obtain
\begin{equation}
  I_+^{\rm tr}=\frac{\gamma_h}{2}p_\uparrow,
  \quad
  I_-^{\rm tr}=0,
  \label{eq:Ipm_lim}
\end{equation}
so the trion EL is completely circularly polarized. Further, from Eq.~\eqref{eq:kinetic} of the main text we find Eq.~\eqref{eq:kin_lim_main} of the main text:
\begin{equation}
  \frac{\d p_{\uparrow/\downarrow}}{\d t}=\pm\gamma_h\left(\frac{1}{2}p_\downarrow-\frac{1}{4}p_\uparrow\right).
  \label{eq:kin_lim}
\end{equation}
These equations can be interpreted as follows. The three exciton states recombine radiatively, while from the fourth dark exciton a singlet trion forms and emits circularly polarized photon, see Fig.~\ref{fig:1photon}(b) in the main text. In this case, an electron is left in the QD with such a spin, that the trion forms with the probability 1/2 after a hole capture by the QD. At the same time, after exciton recombination an electron with random spin is captured, so the probability of trion formation after a hole capture is only 1/4.

\subsection{A. Steady state}

After a single photon emission by a trion, we have $p_\uparrow(0)=1$ and $p_\downarrow(0)=0$. From the solution of Eq.~\eqref{eq:kin_lim} with these initial conditions, we obtain
\begin{equation}
  p_\uparrow(t)=\frac{2}{3}+\frac{1}{3}\e^{-3t\gamma_h/4}.
\end{equation}
Making use of Eq.~\eqref{eq:Ipm_lim} we obtain the second order correlation function at $|\tau|\gg1/\gamma_e$ \addDima{in the steady state:}
\begin{equation}
  g^{(2)}_{s.s.}(\tau)=\frac{p_\uparrow(\tau)}{p_\uparrow(\infty)}=1+\frac{1}{2}\e^{-3|\tau|\gamma_h/4}\addDima{.}
\end{equation}
This expression is plotted in Fig.~\ref{fig:1photon}(d) in the main text. Since $g^{(2)}_{s.s.}(0)>1$, there is a bunching of circularly polarized photons on the time scale $\sim1/\gamma_h$.

\subsection{B. Pulsed excitation}

To model the pulsed electrical excitation of the QD, we assume that there is a constant capture rate of holes $\gamma_h$, while the electron injection with the rate $\gamma_e$ is switched on at time $t=0$ for the time $T$. We assume all the above assumptions for the capture rates to be satisfied, and moreover $\addDima{T}\gg1/\gamma_e$, so at least one electron is already injected.

The photon statistics is described by the probabilities $P_{n}(t)$ of $n$ photon emission at the trion resonance frequency during the time $t$. The second order photon correlation function for a single electrical pulse is defined as
\begin{equation}
  g_p^{(2)}(T)=\frac{\braket{n(n-1)}}{\braket{n}^2},
\end{equation}
where the angular brackets denote the averaging with the probabilities $P_{n}(t)$ in the limit $t\to\infty$.

To calculate $g^{(2)}_p(T)$ we separate the two contributions to it as
\begin{equation}
  P_{n}(t)=P_{n}^\uparrow(t)+P_{n}^\downarrow(t),
  \label{eq:Pnud}
\end{equation}
where the superscript denotes the electron spin state at time $t$. From Eqs.~\eqref{eq:Ipm_lim} and~\eqref{eq:kin_lim} we obtain
\begin{subequations}
  \label{eq:Pn}
  \begin{equation}
    \frac{\d P_{n}^\uparrow}{\d t}=\frac{\gamma_h}{2}P_{n-1}^\uparrow+\frac{\gamma_h}{2}P_{n}^\downarrow-\frac{3}{4}\gamma_hP_{n}^\uparrow,
  \end{equation}
  \begin{equation}
    \frac{\d P_{n}^\downarrow}{\d t}=\frac{\gamma_h}{4}P_{n}^\uparrow-\frac{\gamma_h}{2}P_{n}^\downarrow
  \end{equation}
\end{subequations}
for $0<t<T$. The initial conditions for these equations should be set soon after $t=0$, when the first electron is injected \addDima{into} the QD and the first photon is emitted:
\begin{equation}
  P_{n}^\uparrow(0)=\frac{1}{4}\delta_{n,1}+\frac{3}{8}\delta_{n,0},
  \quad
  P_{n}^\downarrow(0)=\frac{3}{8}\delta_{n,0}.
  \label{eq:init}
\end{equation}
Here we took into account that the resident hole at $t<0$ has no spin polarization.

First of all, from Eqs.~\eqref{eq:Pn} one can see that $\sum_{n=0}^\infty P_{n}$ is constant. It equals to \addDima{one}, in agreement with the initial conditions~\eqref{eq:init}.

Second, we note that
\begin{equation}
  p_{\uparrow/\downarrow}(t)=\sum_{n=0}^\infty P_{n}^{\uparrow/\downarrow}(t).
\end{equation}
Thus, one can check that Eq.~\eqref{eq:kin_lim} follows from Eqs.~\eqref{eq:Pn}. From its solution we obtain
\begin{equation}
  p_\uparrow(t)=\frac{2}{3}-\frac{\e^{-3t\gamma_h/4}}{24},
  \quad
  p_\downarrow(t)=\frac{1}{3}+\frac{\e^{-3t\gamma_h/4}}{24}.
  \label{eq:p_av}
\end{equation}
Here we took into account that $p_\uparrow(0)=5/8$ and $p_\downarrow(0)=3/8$, as follows from the initial conditions~\eqref{eq:init}.

Third, the average number of \addDima{the} emitted photons is given by
\begin{equation}
  \braket{n}=\sum_{n=0}^\infty nP_{n}.
\end{equation}
For the following it is convenient to separate it into two contributions: $\braket{n}=\braket{n}_\uparrow+\braket{n}_\downarrow$,
where
\begin{equation}
  \braket{n}_{\uparrow/\downarrow}(t)=\sum_{n=0}^\infty nP_{n}^{\uparrow/\downarrow}(t)
\end{equation}
are the average numbers of photons emitted at the trion frequency in the cases, when the electron is in the spin up/down state at time $t$. From Eqs.~\eqref{eq:Pn} we obtain
\begin{subequations}
  \begin{equation}
    \frac{\d\braket{n}_\uparrow}{\d t}=\frac{\gamma_h}{2}\braket{n}_\downarrow-\frac{\gamma_h}{4}\braket{n}_\uparrow+\frac{\gamma_h}{2}p_\uparrow,
  \end{equation}
  \begin{equation}
    \frac{\d\braket{n}_\downarrow}{\d t}=\frac{\gamma_h}{4}\braket{n}_\uparrow-\frac{\gamma_h}{2}\braket{n}_\downarrow.
  \end{equation}
  \label{eq:n_av}
\end{subequations}
From the sum of these equations, one can see that
\begin{equation}
  \frac{\d\braket{n}}{\d t}=I_+^{\rm{tr}},
\end{equation}
where $I_+^{\rm{tr}}$ is given by Eq.~\eqref{eq:Ipm_lim}. The solution of this equation with Eq.~\eqref{eq:p_av} reads
\begin{equation}
  \braket{n}=\frac{t\gamma_h}{3}+\frac{2}{9}+\frac{1}{36}\e^{-3t\gamma_h/4}.
\end{equation}
Here we took into account the initial condition $\braket{n}=1/4$ at $t=0$, which follows from Eq.~\eqref{eq:init}. This expression is plotted by the red dashed line in Fig.~\ref{fig:1photon}(e) in the main text.


Fourth, we separate in the same way the two contributions to the second order correlator
\begin{multline}
  \braket{n(n-1)}=\sum_{n=0}^\infty n(n-1)P_{n} \\
  \equiv\braket{n(n-1)}_\uparrow+\braket{n(n-1)}_\downarrow.
\end{multline}
From Eq.~\eqref{eq:Pn} we obtain the kinetic equations for them:
\begin{subequations}
  \begin{multline}
    \frac{\d\braket{n(n-1)}_\uparrow}{\d t}=\frac{\gamma_h}{2}\braket{n(n-1)}_\downarrow\\
    \addDima{-}\frac{\gamma_h}{4}\braket{n(n-1)}_\uparrow+\gamma_h\braket{n}_\uparrow,
  \end{multline}
  \begin{equation}
    \frac{\d\braket{n(n-1)}_\downarrow}{\d t}=\frac{\gamma_h}{4}\braket{n(n-1)}_\uparrow-\frac{\gamma_h}{2}\braket{n(n-1)}_\downarrow.
  \end{equation}
\end{subequations}
We solve these equations together with Eqs.~\eqref{eq:n_av} and~\eqref{eq:p_av} and initial conditions $\braket{n(n-1)}_{\uparrow/\downarrow}=0$, $\braket{n}_\uparrow=1/4$, and $\braket{n}_{\downarrow}=0$ at $t=0$, which follow from Eq.~\eqref{eq:init}. From the solution we obtain
\begin{equation}
  \braket{n(n-1)}=\frac{3(t\gamma_h)^2+8t\gamma_h-2}{27}+\frac{t\gamma_h+8}{108}\e^{-3t\gamma_h/4}.
\end{equation}

Finally, at $t=T$ the electron injection is switched off and only the hole injection is left. After this, no photons can be emitted at the frequency of negatively charged trion, so the second order photon correlation function can be calculated as
\begin{multline}
  g^{(2)}_p(T)=\frac{\braket{n(n-1)}(T)}{\left[\braket{n}(T)\right]^2}\\=
  12\frac{4\left[3(T\gamma_h)^2+8T\gamma_h-2\right]+(T\gamma_h+8)\e^{-3T\gamma_h/4}}{\left(12T\gamma_h+8+\e^{-3T\gamma_h/4}\right)^2}.
\end{multline}
This expression is plotted by the blue solid line in Fig.~\ref{fig:1photon}(e) in the main text. One can check that it tends to one at $T\to\infty$ and to zero at $T\to 0$. It changes at the time scale of the order of $1/\gamma_h$, the slowest time scale in the system.


\end{document}